# Measurement of multiple mechanical properties from multi-dimensional signals in nanosecond laser ablation via PINN

Ying Zhou,[1] Jian Wu,[1, a)] Ziyuan Song,[1] Jinghui Li,[1] Xinyu Guo,[1] Hao Sun,[1] Yuhua Hang,[2] Cuixiang Pei,[3] and Xingwen Li [1]

*[1)] State Key Laboratory of Electrical Insulation and Power Equipment, Xi'an Jiaotong University, Xi'an 710049, China*
*[2)] Suzhou Nuclear Power Research Institute, Suzhou 215004, China*
*[3)] State Key Laboratory for Strength and Vibration of Mechanical Structures, Shaanxi Engineering Research Center of NDT and Structural Integrity Evaluation, Xi'an Jiaotong University, Xi'an, 710049, China*

[a)] Author to whom correspondence should be addressed: jxjawj@mail.xjtu.edu.cn

Accurate evaluation of mechanical properties in steels under ageing or service conditions remains a major challenge. We propose a thermo-mechanical coupling framework for nanosecond laser ablation based on energy conservation, which is embedded into a physics-informed neural network (PINN) to enable simultaneous inversion of multiple mechanical properties. A thermo-mechanical coupling coefficient is defined to uniformly describe the dynamic allocation of input laser energy among thermal diffusion, mechanical work and plasma shielding across different deformation stages under laser irradiation. Furthermore, hard-to-measure physical characteristics in the coupled equation are replaced with experimentally accessible features obtained through the simultaneous acquisition of spectroscopic, shockwave and surface-wave signals. Using 210 experimental datasets, the framework simultaneously recovers Young's modulus, yield strength, ultimate tensile strength and micro-Vickers hardness with high accuracy ($R^2$=0.9927, 0.9912, 0.9916 and 0.9959 respectively), significantly outperforming the baseline method (ultrasonic velocity regression for $E$, $R^2$=0.0012). Comparisons with linear normalization and unconstrained neural networks demonstrate that PINN achieves near-unity accuracy through the embedding of conservation-law constraints. Partial dependency analysis further uncovers the nonlinear coupling laws between input features and mechanical properties. The proposed paradigm, integrating conservation laws, measurable features and physics-informed learning, offers a universal approach for non-contact, high-precision and physically consistent multi-to-multi inversion of multiple material properties under nanosecond laser ablation conditions.

Accurate evaluation of mechanical properties is critical for steels, particularly in aging or in-service components where safety is paramount[1]. Conventional destructive methods, such as uniaxial tensile testing, provide direct stress-strain curves but are unsuitable for service monitoring. A variety of non-destructive techniques have been investigated: X-ray diffraction probes residual stress and dislocation density near surfaces[2], while neutron diffraction extends to bulk but requires large-scale facilities and offers limited resolution[3]. Raman/IR spectroscopy detects lattice strain but lacks robust mapping to strength[4]. Nanoindentation measures hardness and modulus but is constrained by locality and size effects[5]. Digital image correlation provides full-field strain but requires external loading and surface preparation[6]. Overall, these approaches remain confined to single-parameter analysis, limiting their capacity to evaluate steels during ageing.

Laser-based methods offer promising alternatives[7,8]. Laser-induced breakdown spectroscopy (LIBS) has been employed to assess aging and hardness from spectral features[9], yet correlations remain largely empirical and sensitive to matrix effects. Laser ultrasonics enables non-contact, bulk probing under harsh conditions[10]; nevertheless, conventional time-of-flight signals suffer from poor sensitivity[11] due to multimode aliasing[12,13] and primarily reflect elastic modulus[14-16]. Nonlinear ultrasonics broadens inversion capability[17,18], but its parameters are affected by diverse defects, leading to non-unique interpretation[19,20]. Moreover, weak harmonic signals are readily contaminated by noise[21].

Unlike single-parameter techniques, laser-target interaction in ablation mode inherently excites multiple responses[22], including plasma emission, shockwave and acoustic transients[23,24]. These strong, information-rich signals hold potential for more comprehensive evaluation[25,26]; nonetheless, complex coupling, condition sensitivity and reliance on empirical regression still impede systematic and quantitative application[21,27,28].

To address this issue, this letter focuses on steels as representative structural materials, deriving thermo-mechanical coupling relations under laser ablation from energy conservation and embedding them into a physics-informed neural network (PINN)[29,30] for quantitative inversion of multiple mechanical properties. Physical interpretability of this framework is further examined through partial dependency analysis.

To emulate service ageing, #45 carbon structural steel was subjected to heat treatment to generate specimens with distinct mechanical states. Samples were heated, held and cooled in vacuum for 1, 3, 5, 7 and 9 hours, with the holding temperature maintained just below the ferrite-to-austenite transformation threshold of medium carbon steel. This protocol was sufficient to produce clear and reproducible variations in mechanical parameters (details in Ref. 31).

Theoretical analysis begins by establishing a thermo-mechanical coupling framework based on the first law of thermodynamics. For a micro-control volume within the target under nanosecond laser irradiation, the absorbed energy is partly dissipated as heat and partly converted into mechanical work. The entropy production $\rho T\dot{s}$ can thus be expressed as

$$I_{\text{target}} - \nabla \cdot q = \rho T\dot{s} = \rho c_{\varepsilon} \frac{\partial T}{\partial t} - T\frac{\partial \sigma}{\partial T}\frac{\partial \varepsilon}{\partial t}, \qquad (1)$$

where $I_{\text{target}}$ is the absorbed laser energy, $\nabla \cdot q$ is the net thermal flux, $T$ is the target temperature, $c_{\varepsilon}$ represents specific heat at constant strain, $\sigma$ is stress and $\varepsilon$ is strain.

The term $-\partial\sigma/\partial T$ in Eq.(1), defined as the thermo-mechanical coupling coefficient $\beta$, can be expanded by decomposing strain into thermal and mechanical contributions,

$$-\frac{\partial\sigma}{\partial T}=-\left(\frac{\partial\hat{\sigma}}{\partial\varepsilon^{m}}\frac{\partial\varepsilon^{m}}{\partial T}+\frac{\partial\hat{\sigma}}{\partial T}\bigg|_{\varepsilon^{m}}\right)=\alpha H_{t}-\frac{\partial\hat{\sigma}}{\partial T}\bigg|_{\varepsilon^{m}}\triangleq\beta, \quad (2)$$

where $\alpha$ is the thermal expansion coefficient, $H_t$ is isothermal tangent modulus. Substituting Eq.(2) into Eq.(1) yields the complete thermo-mechanical coupling equation

$$\nabla\cdot(-k\nabla T)+\rho c_{\varepsilon}\frac{\partial T}{\partial t}+T\underbrace{\left(\alpha H_{t}-\frac{\partial\hat{\sigma}}{\partial T}\right)}_{\beta}\frac{\partial\varepsilon}{\partial t}=I_{\text{target}}, \quad (3)$$

Steels exhibit distinct yield plateau and strain hardening versus alloys/ceramics, making their laser ablation response particularly revealing. Phase transitions occur in three stages: (i) optical absorption: photon-induced lattice vibrations reduce dislocation activation energy, triggering dislocation slip/annihilation and initiating softening by lowering ultimate tensile strength UTS and micro-Vickers $H_v$ via Taylor hardening[32,33]; (ii) thermal diffusion: Gibbs-Thomson-driven grain coarsening (Ostwald ripening) modulates elastic modulus $E$ through boundary density evolution: initial coalescence elevates $E$ via pseudo Hall-Petch effects, whereas excessive coarsening activates boundary sliding[34-36]; (iii) phase explosion: plasma confinement induces stress-assisted void nucleation (electron number density ($n_e$)-dependent mass flux), degrading yield strength $\sigma_s$ via Griffith fracture initiation[37,38].

Concurrently, deformation progresses through distinct regimes. Initially, elastic energy storage dominates ($H_t \approx E$ and $\beta \approx \alpha E$). As thermal softening intensifies, the material enters the yield plateau ($\beta \approx -\partial\sigma_s/\partial T$), where plastic flow initiates permanent deformation. Subsequent plastic hardening ties $\beta$ to the hardening modulus $H_p$ and temperature sensitivity of flow stress ($\partial\hat{\sigma}/\partial T|_{\varepsilon m}$), with $H_t=EH_p/(E+H_p)$. Upon reaching UTS ($H_t \to 0$), necking localizes deformation, collapsing both $H_t$ and true stress $\sigma_{\text{true}}$ as energy allocation shifts entirely to phase transitions (solid, liquid, to plasma states), completing the ablation lifecycle.

From the above analysis, it follows that during various stages of laser-target interaction, the partitioning of energy among thermal diffusion, mechanical work and phase transition is dynamically regulated by $\beta$. According to Eq.(3), monitoring $T$, $\varepsilon$ and $I_{\text{target}}$ would, in principle, allow simultaneous inversion of multiple mechanical properties. Because these parameters are difficult to access directly, they are instead mapped onto measurable proxies for practical implementation.

Specifically, target surface temperature $T_s$ can be approximated by vapor temperature at the liquid-gas interface under the half-range Maxwellian of the surface evaporation model[39], further estimated by electron temperature $T_e$ for local thermodynamic equilibrium. $\varepsilon$ can be inferred from strain energy $E_s$ associated with surface vibration: $\varepsilon \propto \sqrt{E_s}$ for elastic deformation, and $\varepsilon \propto E_s$ for yielding and plastic deformation.

The laser energy reaching the target, $I_{\text{target}}$, is derived from the incident energy $I_0$ according to Beer-Lambert attenuation and modified by plasma shielding

$$I_{\text{target}}=I_{0}\exp\left(-\left(\kappa_{IB}(T_{e},n_{e},N)+\kappa_{PI}(T_{e},n_{e},N)\right)L\right), \quad (4)$$

where $\kappa_{\text{IB}}$ and $\kappa_{\text{PI}}$ are radiation absorption coefficients for inverse bremsstrahlung and photoionization, respectively, both governed by $T_e$, $n_e$, and total particle number density $N$ within the plasma. $L$ is the plasma length.

Spectral intensity follows radiation transport equation for optically thick plasma[40]: $I_{\text{spec}}=B[1-\exp(-\kappa_{\text{line}}L)]$, here $B$ is Planck function determined by $T_e$. $\kappa_{\text{line}}$ is the line absorption coefficient, further expanded

$$\kappa_{\text{line}}=\frac{\pi e^{2}f}{m_{e}c}g_{\text{low}}n_{i}\frac{e^{-E_{\text{low}}/k_{B}T_{e}}}{Z_{i}(T_{e})}\left(1-e^{-h\nu/k_{B}T_{e}}\right)P(\nu,n_{e}), \quad (5)$$

where $Z_i$ is partition function, $n_i$ is the ion number density in the $i$-th ionization state, $P(\nu, n_e)$ is the line profile function determined by $n_e$ under Stark broadening. Consequently, $I_{\text{spec}}$ depends on $T_e$, $n_e$, $n_i$ and $L$. In practice, $T_e$ can be obtained from the Saha-Boltzmann plot, $n_e$ from spectral line FWHM, and $I_{\text{spec}}$ directly from emission measurements, enabling simultaneous acquisition of $T_e$, $n_e$ and $I_{\text{spec}}$.

In addition, shockwave pressure during plasma expansion is monitored to complement estimation of $N$, with $P=Nk_BT_e$ under the ideal gas assumption.

Accordingly, Eq.(3) can be reformulated into a constraint involving measurable variables ($T_e$, $n_e$, $I_{\text{spec}}$, $P$, $E_s$)

$$H(T_{e})+G(\beta,E_{s})=I_{0}-F(T_{e},n_{e},I_{\text{spec}},P), \quad (6)$$

PINN is then developed for mechanical property inversion. The network adopts a shallow, fully connected feedforward structure to minimize overfitting while exploiting strong constraints of physical equations[41]. It consists of an input, two hidden (eight neurons each), and an output layer, with Tanh activations introducing nonlinearity. The input vector is $X=(T_e, n_e, I_{\text{spec}}, P, E_s)$ and the output is $Y=(E, \sigma_s, \text{UTS}, H_v)$, where $Y$ represents mechanical responses during elastic deformation, yielding and failure under laser irradiation. All variables were standardized to maintain physical multiplicative relationships.

The forward propagation is expressed as

$$Y_{\text{pre}}=\tanh\left(\tanh\left(XW_{1}^{T}+b_{1}\right)W_{2}^{T}+b_{2}\right)W_{3}^{T}+b_{3}, \quad (7)$$

where $W_i$ and $b_i$ ($i$=1,2,3) are weights and biases of each layer.

Unlike black-box regressors, PINN minimizes the total loss $\mathscr{L}_{\text{total}}=\mathscr{L}_{\text{data}}+\lambda\cdot\mathscr{L}_{\text{phys}}$. Physical weight $\lambda$ balances data fidelity and physical consistency, physical loss $\mathscr{L}_{\text{phys}}$ quantifies deviation between the left and right sides of Eq.(6). Specific expressions of data loss $\mathscr{L}_{\text{data}}$ and $\mathscr{L}_{\text{phys}}$ are given

$$\begin{cases}\mathscr{L}_{\text{data}}=MSE(Y-Y_{\text{pre}})=\frac{1}{4}\sum_{i=1}^{4}\left\|Y^{i}-Y_{\text{pre}}^{i}\right\|_{2}^{2}\\ \mathscr{L}_{\text{phys}}=MSE\left(I_{0}-F(I_{\text{spec}},T_{e},n_{e},P)-G(\beta,E_{s})-H(T_{e})\right)\end{cases}, \quad (8)$$

The following details experimental configuration for multi-signal acquisition (Figure 1). An Nd:YAG pump-beam (Lapa 80, BeamTech) with 1064 nm wavelength, 10 ns pulse width, 3 Hz repetition rate, and ~20 mJ energy was split by a beam-splitter (BSF10-C, Thorlabs). The transmitted fraction was sampled by a photodiode (DET10A, Thorlabs) for timing, while the reflected laser was focused onto the sample surface at 45° using a convex lens (f=30 mm) to generate plasma with ~80 μm spot diameter. Plasma emission was collected by a short-focal lens and transmitted via an optical fiber to an echelle spectrometer (LTB, Aryelle Butterfly, 270-690 nm spectral range, ~12,500 resolution, Germany) with 600 ns gate delay and 10 μs gate width. A digital delay pulse generator (Stanford Research Systems, DG645, USA) synchronized the laser and spectrometer.

A 532 nm wavelength, ~1.5 W power probe-beam from a laser interferometer (Quartet, Sound & Bright) was normally focused (f=100 mm) to monitor out-of-plane displacement of the surface. The pump-probe distance was set to ~5 mm to avoid ablation craters. A pressure sensor (113B26, PCB Piezotronics, Inc.) was mounted at 30° and ~10 mm distance from ablation regions to measure shockwave. An oscilloscope (Tektronix MDO-3034, 500 MS/s) simultaneously acquired laser excitation, shockwave and surface wave signals, averaging 128 acquisitions to suppress environmental noise. Five random sites were selected on each sample surface, each cleaned with 20 pre-ablation before acquiring 150 cumulative shots for each spectrum, shockwave and surface wave.

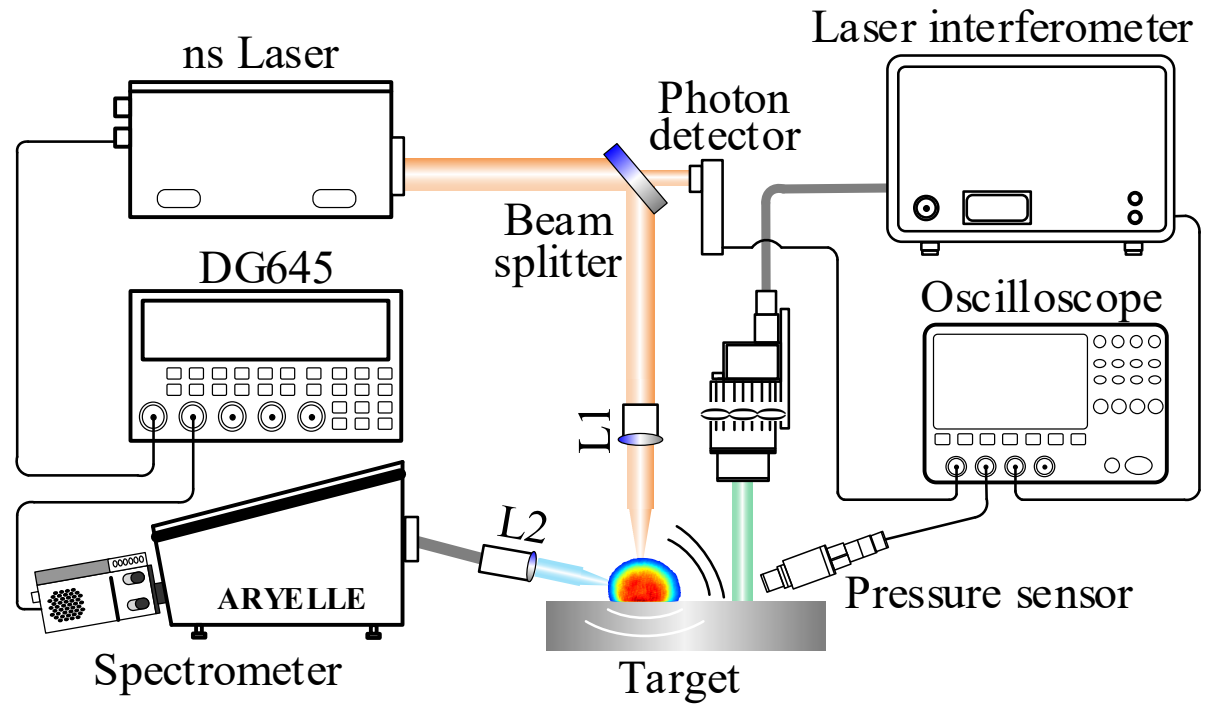

**FIG. 1** Experimental configuration for multi-signal acquisition.

Complementary mechanical testing was performed on heat-treated samples. Uniaxial tensile tests yielded stress-strain curves, from which $E$ was obtained and $\sigma_s$ determined using the 0.2% offset method. UTS was identified as the maximum stress, while $H_v$ was measured with a micro-Vickers hardness tester. (Further detailed are provided in Ref. 31)

Figure 2(a) presents mechanical testing results for the blank and five heat-treated samples. The error lines originate from 16 points, which are randomly selected on both sides of the sample. Prolonged annealing causes gradual decreases in $\sigma_s$, UTS and $H_v$, consistent with the decline in dislocation density estimated from X-ray diffraction using modified Williamson-Hall method[31]. In contrast, $E$ changes only slightly, rising from 192.9 GPa to 195.4 GPa at 3 h before gradually returning to 192.7 GPa at 9 h. Figure 2(b) shows time-domain waveforms of surface wave and shockwave for the blank. A 2~8 MHz bandpass filter is applied to the original surface wave signal to match its center frequency and suppress noise. The first surface wave with a central, bipolar feature appears ~2 μs after laser excitation, followed by a second echo at ~9 μs. The shockwave signal is processed with a 500 kHz low-pass filter to align with the resonant frequency of the pressure sensor. Figure 2(c) displays the emission spectrum of the blank, which is dominated by atomic and ionic lines of matrix iron due to the low alloy content of #45 carbon steel.

Figure 3 summarizes the evolution of four physical characteristics with annealing time. Surface wave velocity $v_s$ is calculated from the first surface wave arrival and pump-probe distance. Surface wave energy $E_s$ is determined by squaring the first surface amplitude and integrating over time (Figure 3(a)). Shockwave peak pressure $P$ is extracted and the shockwave energy $E_p$ is determined from the squared peak profile integrated over time (Figure 3(b)). Figure 3(c) shows net peak intensities of two representative Fe I lines. $T_e$ is derived using the Saha-Boltzmann plot with five Fe II lines and seven Fe I lines from the locally-magnified spectrum in Figure 2(c). $n_e$ is calculated using FWHM of Fe I 404.58 nm line (Figure 3(d)). The consistency evolution of $v_s$ and $E$ corroborates the relationship $v_s \propto \sqrt{E_s/\rho}$. In addition, the net peak intensities of Fe I lines decreases continuously with annealing duration.

More complex evolution appears in other characteristics. During 0~3 h period (Stage I), thermal activation reduces dislocation density and dissolves cementite phase, collectively weakening lattice slip resistance[42,43]. The dissolution kinetics, governed by an Arrhenius-type relation between thermal driving force (which can be characterized by $T_e$ to some extent) and dissolution rate , drives continuous declines in $\sigma_s$/UTS/$H_v$[44]. Through Eq.(3), this mechanical softening lowers $\beta$ and progressively diminishes $E_s$. Target softening decreases $P$, in agreement with the continuous decline of $E_p$ and $I_{\text{spec}}$ (as parts of the plasma internal energy), and the slight drop in $T_e$. Enhanced thermal absorption elevates evaporation mass, driving persistent $n_e$ growth. In the subsequent 3~9 h period (Stage II), grain coarsening exacerbates grain boundary weakening, forcing further $\beta$ reduction as $\sigma_s$/UTS/$H_v$ keep declining[45,46]. Plasma shielding fluctuation induces non-uniform heat input, broadening the grain size distribution and corresponding to the continued UTS decrease. $E_p$ attenuation reflects shielding weakening, temporarily raising $I_{\text{target}}$ and inducing $E_s$ recovery (5 h). However, mechanical degradation ultimately drives $E_s$ down again (7~9 h). The interplay of these factors raises the energy absorbed as heat, increasing $T_s$ and $T_e$, which not only slows $I_{\text{spec}}$ decay but also reactivates $P$ and $E_p$ as plasma becomes hotter. Restored shielding temporarily suppresses evaporation efficiency, causing $n_e$ dip at 5 h. This dynamic competition drives Stage II to evolve differently from Stage I, underscoring the complex relationship between annealing-driven microstructural evolution and coupling of energy allocation in laser ablation.

The evolution of aforementioned physical characteristics alongside mechanical properties confirms the necessity of energy conservation. Figure 4 demonstrates inversion

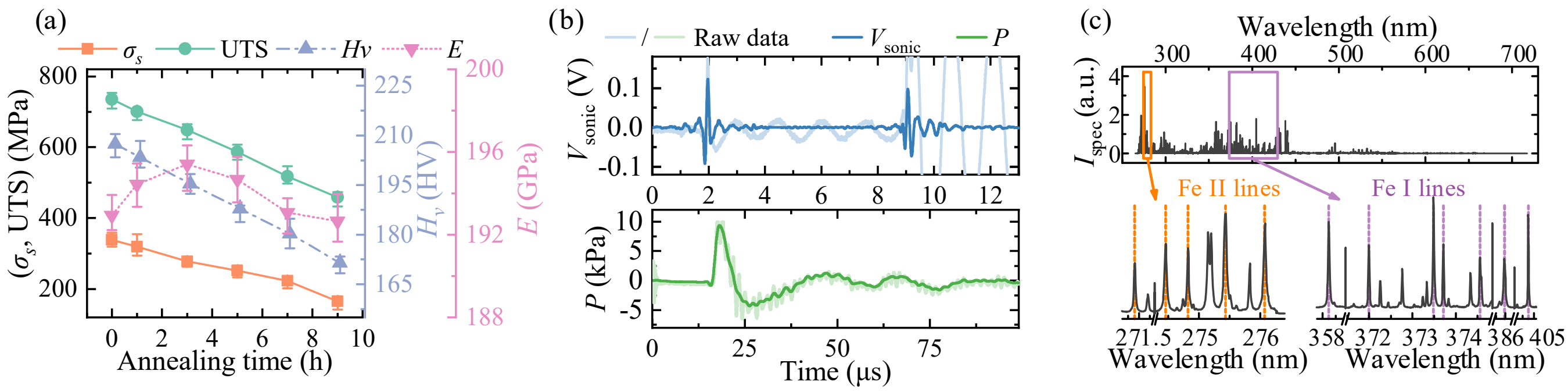

**FIG. 2** (a) Samples' mechanical properties for different annealing time. (b)Surface wave, shockwave and (c)spectrum for the untreated sample.

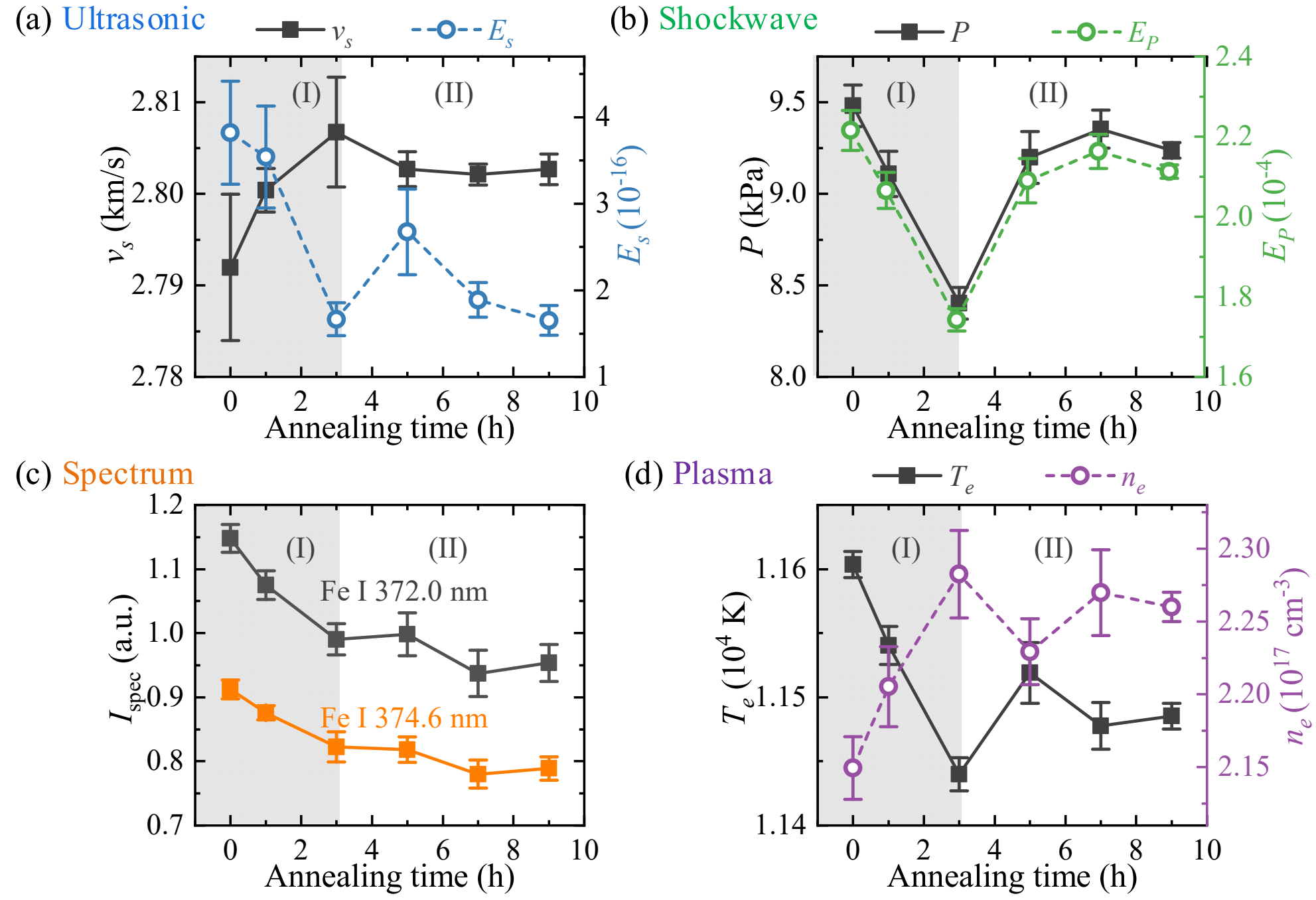

**FIG. 3** Evolution of physical features with annealing time: (a)surface wave; (b)shockwave; (c)spectrum; (d)plasma characteristics.

performance based on this conservation. To expand the training dataset and mitigate overfitting, $I_{\text{spec}}$ in vector $X$ is taken as a collection of seven Fe I lines (marked in Figure 2(c)), yielding 5×7=35 data points per sample and 210 in total. Network uses an 80:20 split for training and validation sets. The Adam optimizer is employed with 0.001 learning rate to balance convergence speed and stability. $\lambda$ is set to 0.1875 to enforce physical constraints. Batch size is set to 32 balancing the efficiency of mini-batch gradient descent with memory overhead, and training epochs are set to $10^4$. Figure 4(a) shows monotonic decreases and subsequent stabilization of loss functions after $3\times10^3$~$10^4$ epochs. The rapid decline of $\mathscr{L}_{\text{phys}}$ in the early training stage indicates that physical laws were effectively captured. By enforcing energy constraints, PINN converges toward physically admissible solutions rather than relying solely on statistical correlations. Subsequently, $\mathscr{L}_{\text{data}}$ and validation loss $\mathscr{L}_{\text{val}}$ converge smoothly, demonstrating that the model effectively balances data fitting with physical consistency.

Figure 4(c)-(f) illustrate regression performances of PINN for four mechanical properties. Compared to direct regression based on $v_s$ (Figure 4(b)), the correlation between $v_s$ and $E$ is negligible ($R^2$=0.0012). This is because during thermal ablation, $v_s$ is influenced not only by $E$, but also by coupling effects of thermo-mechanical softening and plasma shielding. It renders the simple mapping between the two ineffective. By contrast,

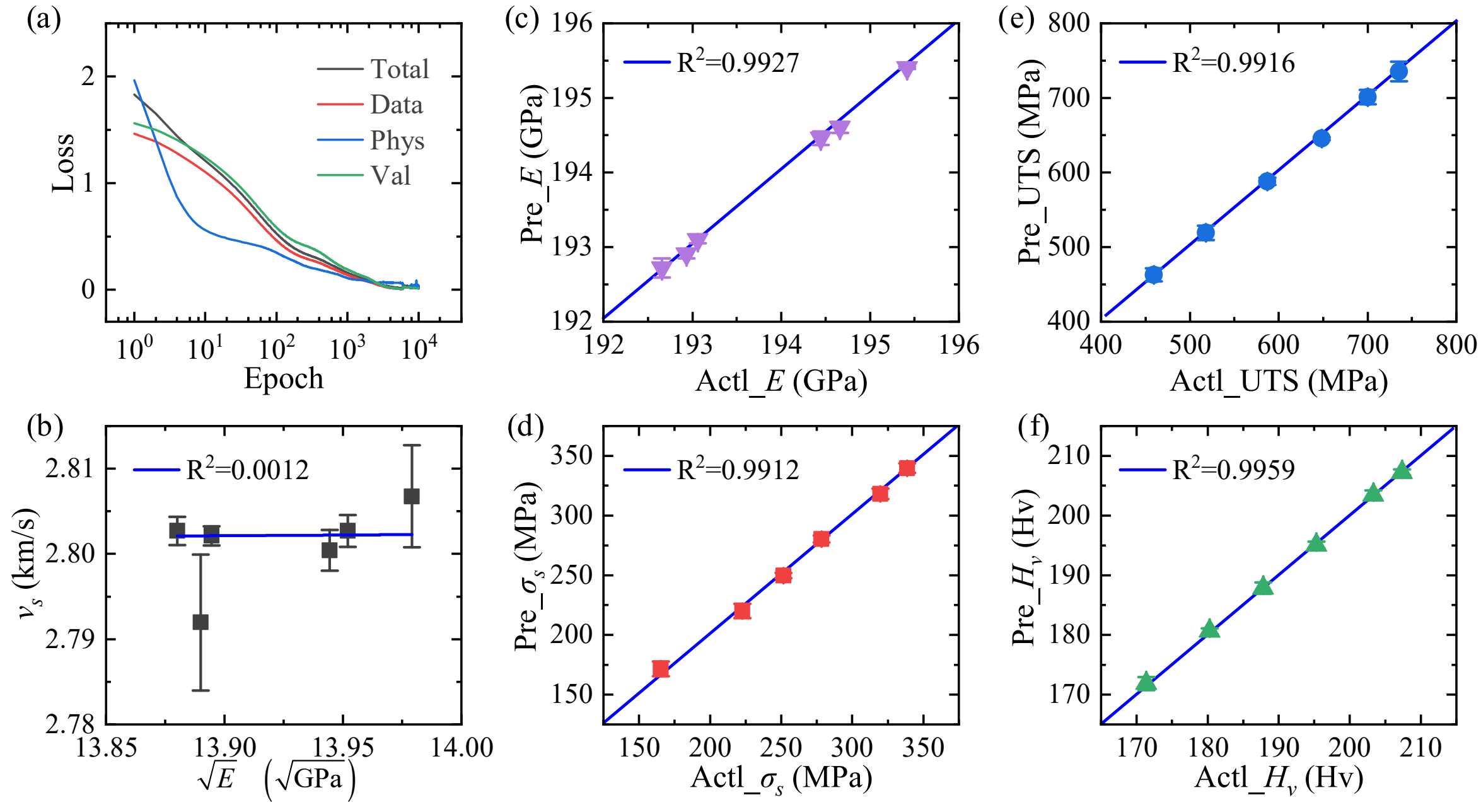

**FIG. 4** (a) PINN's loss functions. (b) Correlation between $v_s$ and $E$. Regression performances of PINN for (c)$E$, (d)$\sigma_s$, (e)UTS and (f)$H_v$.

PINN exhibits excellent accuracies for $E$, $\sigma_s$, UTS and $H_v$ ($R^2$=0.9927, 0.9912, 0.9916 and 0.9959). This significant improvement stems from embedding physical constraints into learning, enabling the network to capture nonlinear laser-target interactions beyond empirical fitting.

A more detailed analysis follows, with Figure 5 illustrating comparative performance and interpretation. Figure 5(a) compares regression performance using linear normalization (LN), feedforward neural network without physical constraints (FNN) and PINN. For each mechanical property in LN, the single input feature that yields the best performance in regression is selected. LN yields only moderate accuracy and performs poorly for $E$ ($R^2$=0.6495). FNN, which incorporates all five features, improves prediction ($R^2$ lies between 0.95 and 0.98), but remains data-driven. By embedding conservation-law constraints into the same architecture, PINN achieves near-unity accuracy across all properties. These results highlight how physical constraints enhance predictive power while ensuring consistency with underlying thermo-mechanical mechanisms, making PINN well-suited for reliable multi-property inversion.

Figure 5(b)-(e) present partial dependence plots of PINN outputs versus input features. Each mechanical property exhibits distinct coupling with input physical quantities. $T_e$, $n_e$ and $P$ shows strong positive or negative correlations with strength-related properties ($\sigma_s$, UTS and $H_v$), but weaker sensitivity to $E$, reflecting the short duration of elastic deformation and the dominance of yield and subsequent stages during thermal ablation and phase transitions. $P$ exhibits a phased response: initial insensitivity to strength-related properties (negative half-cycle), followed by a sharp response (positive half-cycle). This suggests dual control by total particle number density (derived from ablation mass) and plasma shielding. The steep slope may also be partly due to target hardening, which amplifies the recoil generated at the surface during plasma expansion. $E_s$ exhibits a unimodal response, revealing the dynamic allocation of $I_{target}$ between heat diffusion and mechanical work. $I_{spec}$ correlates almost linearly with all mechanical properties, partially reflecting the plasma internal energy (the dependency of $I_{spec}$ is significantly diluted by expansion into a 7-bar set within $X$).

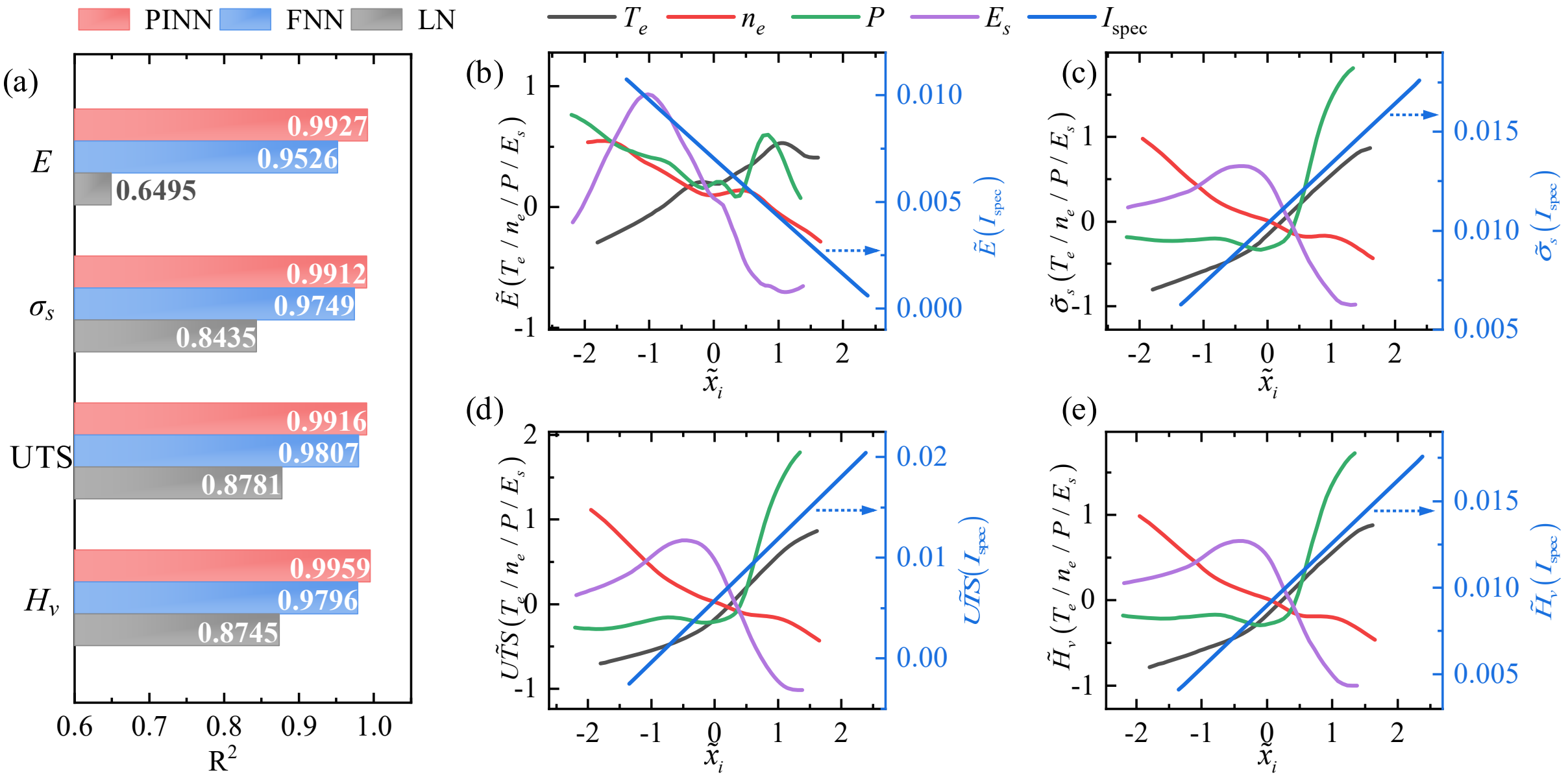

**FIG. 5** (a)Regression performance for LN, FNN and PINN. Partial dependence plots of PINN outputs, (b)$E$, (c)$\sigma_s$, (d)UTS, (e)$H_v$, on input features.

In summary, this study establishes a thermo-mechanical coupling control equation for laser-target interaction in thermal ablation mode based on the principle of energy conservation. The introduction of thermo-mechanical coupling coefficient $\beta$, incorporates mechanical properties across successive deformation stages under laser irradiation, clarifying their regulatory role in dynamically allocating energy among heat diffusion, mechanical work and plasma shielding. Within this framework, hard-to-measure observables in constraint equations are replaced by accessible quantities: plasma emission, shockwave and surface wave ($T_e$, $n_e$, $I_{spec}$, $P$, $E_s$). PINN is then constructed to invert four mechanical properties ($E$, $\sigma_s$, UTS, $H_v$) simultaneously. Compared with direct regressing $E$ from $v_s$ ($R^2$=0.0012), the proposed inversion strategy markedly improves regression accuracy for all four mechanical properties ($R^2$=0.9927 for $E$, 0.9912 for $\sigma_s$, 0.9916 for UTS and 0.9959 for $H_v$). Importantly, comparative analyses with LN and unconstrained FNN confirm that while LN yields only moderate accuracy and FNN improves prediction yet remains data-driven, PINN achieves near-unity accuracy by embedding conservation-law constraints. Partial dependence analysis further reveals nonlinear relationships between input features and output mechanical properties in thermal ablation mode, demonstrating a physically consistent multi-to-multi inversion strategy with clear interpretability. Future work will extend the PINN framework to complex service environments by advancing its architecture through deep learning and integrating advanced nonlinear ultrasonic parameters to improve characterization of dislocation structures and micro-damage.

## SUPPLEMENTARY MATERIAL

See the supplementary material for sensitivity analysis of physical weight $\lambda$ and validation on creep-damaged 800H

superalloy underscores our work's potential for in-service monitoring applications.

## ACKNOWLEDGMENTS

This work was supported by the National Key Research and Development Program of China (Grant No. 2022YFB3707200) and Shaanxi Provincial Science and Technology Plan Project (2024TC-DXWT-04). The authors acknowledge Dr. I.B. Gornushkin from BAM Federal Institute for Materials Research and Testing for many inspiring discussions about the model.

## AUTHOR DECLARATIONS

### Conflict of Interest

The authors have no conflicts to disclose.

## DATA AVAILABILITY

The data that support the findings of this study are available from the corresponding author upon reasonable request.